\documentclass[lettersize,journal]{IEEEtran}
\usepackage{amsmath,amsfonts}
\usepackage{algorithmic}
\usepackage{algorithm}
\usepackage{array}
\usepackage[caption=false,font=normalsize,labelfont=sf,textfont=sf]{subfig}
\usepackage{textcomp}
\usepackage{stfloats}
\usepackage{url}
\usepackage{verbatim}
\usepackage{graphicx}
\usepackage{cite}
\usepackage{booktabs} 
\usepackage{tabularx} 
 \usepackage{ragged2e} 
 \usepackage{enumitem} 

\newcolumntype{L}{>{\RaggedRight\arraybackslash}X}

\begin{document}

\title{Hands-On Beyond Classrooms: Elevating Control Engineering Education with Take-Home Labs}


\author{%
\IEEEauthorblockN{Martin Goubej\IEEEauthorrefmark{1},~\IEEEmembership{Member,~IEEE},
John Hedengren\IEEEauthorrefmark{2}, ~\IEEEmembership{Member,~IEEE}, Daniel Sukovat\'{y}\IEEEauthorrefmark{1}\\}
\IEEEauthorblockA{\IEEEauthorrefmark{1}NTIS Research Centre, University of West Bohemia,
Pilsen, Czechia\\
Email: \{mgoubej,dansuk24\}@fav.zcu.cz\\
}
\IEEEauthorblockA{\IEEEauthorrefmark{2}Department of Chemical Engineering, Brigham Young University, Provo, UT, USA\\
Email: john\_hedengren@byu.edu}
}




\maketitle

\begin{abstract}
\emph{Contribution:} This study is positioned within the Scholarship of Application, showing how take-home laboratories can be integrated into control engineering education to strengthen hands-on learning in both introductory and advanced courses. The work provides evidence that a compact experimental platform can support conceptual understanding, student engagement, and course performance in a broadly applicable way.

\emph{Background:} Control engineering courses often rely heavily on theory and simulation, which may limit students’ ability to connect abstract concepts with practical implementation. Low-cost take-home laboratories offer a scalable alternative that can help bridge this gap across different educational settings.

\emph{Intended Outcomes:} The curricular change, consisting of the introduction of take-home laboratory kits into the semester projects, was intended to improve students’ motivation, deepen their understanding of core control principles, and enhance their performance in coursework and final assessment.

\emph{Application Design:} Students used the kits for modeling, identification, and control design tasks, while learning outcomes were evaluated through questionnaires, midterm test scores, and final exam grades.

\emph{Finding:} Student feedback indicated strong acceptance of the new format and improved understanding of the subject matter. Statistical analysis showed a significant difference between cohorts before and after the curricular change, with modest but measurable effect sizes. The results suggest that take-home laboratories can support both motivation and academic performance, while also improving students’ appreciation of the practical relevance of control engineering.

\end{abstract}

\begin{IEEEkeywords}
Control engineering education, Take-home laboratories, Project-based learning, Hands-on learning, Single and multivariable control, Robust control, PID control
\end{IEEEkeywords}

\section{Introduction}

\IEEEPARstart{T}he importance of hands-on laboratory experiences in control engineering education has been widely recognized \cite{Comault82,Craig2001,Schiller13,Wang24}. Traditional laboratory environments, while effective, often require expensive infrastructure and significant logistical efforts. As a response, various simplified problem-oriented setups \cite{quanser,goubej2022raspberry}, hardware-in-the-loop simulators \cite{SOBOTA2019_ace} or portable and take-home laboratory kits, such as the Temperature Control Laboratory (TCLab), have emerged as a cost-effective and accessible alternative \cite{wang_toe_motor,oliveira2023tclab, hedengren2020tclab}.

Several studies have demonstrated the effectiveness of TCLab in teaching fundamental concepts in process control and system dynamics. Oliveira et al. \cite{oliveira2023tclab} discuss the adoption of TCLab in Portugal to highlight the potential to bridge theoretical topics with practical implementation. Similarly, Hedengren \cite{hedengren2020tclab} highlights how TCLab can be integrated into undergraduate courses to provide students with an interactive platform to explore feedback control principles.

Recent advances in artificial intelligence and reinforcement learning have also influenced control education. Patel \cite{patel2023rl} explores how reinforcement learning can be implemented in continuous process control, potentially enhancing adaptive control strategies in educational setups. Additionally, Lee and Kang \cite{lee2022deeptclab} compare deep learning models for temperature regulation to emphasize the role of machine learning in process control.

Portable laboratory setups also contribute to the advancement of modern control strategies. Mejía et al. \cite{mejia2022smithpredictor} evaluate various Smith predictor approaches for thermal processes with significant dead time to illustrate how experimental platforms like TCLab can facilitate comparative control studies. Similarly, Yerolla and Besta \cite{yerolla2021pid} introduce tuning-free PID controllers tailored for systems with time delays to demonstrate an application that aligns with TCLab-based control projects.

Furthermore, the pedagogical impact of take-home labs has been assessed in the context of engineering education. Moura Oliveira et al. \cite{oliveira2022portablelabs} present pocket-sized portable labs as an effective tool for control engineering practice to reinforce the idea that hands-on experiences can enhance conceptual understanding. 



In this work, we embed the take-home laboratory kits into a project-based learning (PjBL) and competency-based education (CBE) framework, in which students progress through a sequence of authentic engineering tasks and are assessed against explicitly stated competencies. Within this framework, the main goal of the paper is to exploit the versatility of the kit in a more complex multiple-input multiple-output (MIMO) setup, allowing students to acquire practical skills that are relevant for many real-world applications that are inherently multivariable, ranging from distillation columns and chemical reactors to aircraft flight control systems, smart grids, or robotic systems \cite{skogestad2005multivariable}. 

NOTE: THIS IS AN AUTHORS' PREPRINT VERSION. THIS WORK HAS BEEN SUBMITTED TO THE IEEE FOR POSSIBLE PUBLICATION. COPYRIGHT MAY BE TRANSFERRED WITHOUT NOTICE, AFTER WHICH THIS VERSION MAY NO LONGER BE ACCESSIBLE.

\newpage

The project is organized as a gradual progression through several learning modules, starting with first-principles modeling and data-driven identification, continuing with classical control design methods, and culminating in advanced multivariable control concepts, thereby guiding students through a~complete model-based control engineering cycle encountered in practice. Each module is aligned with a set of expected learning outcomes (ELOs) that map onto five broader control engineering competency domains, supporting constructive alignment between course objectives, learning activities, and assessment. A secondary goal of the paper is to evaluate students’ feedback and their academic performance, as measured by midterm test scores and final exam grades, in order to assess the educational impact of the proposed PjBL–CBE design.

To guide the study, we formulate the following research questions:
\begin{enumerate}
\item[RQ1:] How does the introduction of take-home TCLab projects affect students’ perceived understanding of control engineering concepts and their engagement with the course?
\item[RQ2:] To what extent does the project-based, competency-aligned curriculum improve students’ performance in midterm tests and final exam grades compared to previous cohorts?
\item[RQ3:] How do the proposed learning modules support the development of competencies across the five targeted control engineering domains?
\end{enumerate}

The paper is structured as follows. Section II briefly introduces the TCLab kits from the hardware and software perspective. Section III presents a structure and content of the control engineering courses, where the TCLab kits were employed recently. Sections IV and V provide a study on integration of the take-home laboratories to the existing study programs and course curricula, presenting the ELOs of each learning module. Finally, Section VI deals with the mapping of the targeted competency domains with related learning outcomes, student activities and evidences of achievement, followed by a validation study measuring the improvements in students' academic performance. Concluding remarks are given in Section VII.

\section{TCLab Take-home kits}

\begin{figure}[!t]
\centering
\includegraphics[width=0.485\textwidth]{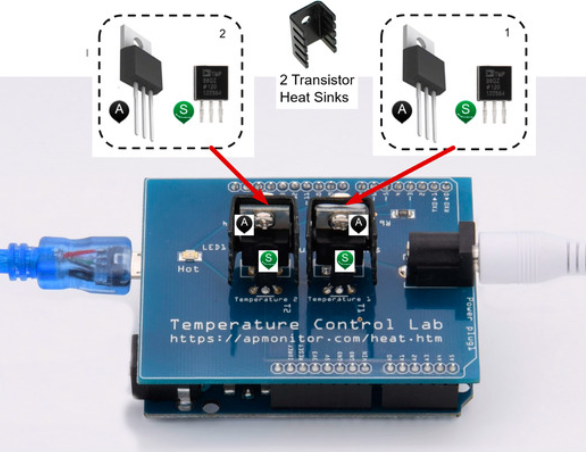}
\caption{Temperature Control Laboratory - Arduino-based electronic board with two heater elements serving as actuators (A) and two temperature sensors (S) constituting a pocket-sized take-home kit}
\label{fig_2}
\end{figure}

The Temperature Control Laboratory (TCLab) is a compact, Arduino-based experimental platform designed for hands-on learning in process control (Fig. \ref{fig_2}). The hardware consists of an Arduino Leonardo or compatible microcontroller, two transistor-based heaters, and two TMP36 analog temperature sensors for real-time thermal measurements. The system is powered with a USB or an external power adapter, with heater power levels adjustable between 0-100\% with Pulse-Width Modulation. The device communicates through a serial USB interface to integrate with various control and data acquisition systems.

TCLab is equipped with software interfaces for MATLAB and Python for flexible implementation of control algorithms. In MATLAB, the \texttt{tclab.m} package provides functions for heater control, temperature reading, and real-time data logging. MATLAB users must install the Arduino Support Package to enable communication with the microcontroller. Similarly, Python users utilize the \texttt{tclab.py} package (\texttt{pip install tclab}), to facilitate interaction with the hardware through serial commands. The Python interface supports real-time plotting, PID tuning, and system identification through libraries like \texttt{Matplotlib}, \texttt{NumPy}, and \texttt{SciPy}. Additionally, the \texttt{TCLabModel()} function in Python gives offline simulation of the system for cases where hardware is unavailable.

Browser-based applications provide an additional entry point to TCLab learning activities without requiring a local MATLAB or Python installation. The TCLab Control Studio supports interactive experiments with a simulated device and direct operation of a physical TCLab through a compatible web browser. With hardware connected, a digital twin runs alongside the physical device, allowing students to compare predicted and measured responses. The additional TCLab Simulation Studio enables students to simulate thermal dynamics, estimate model parameters from simulated or experimental data, and compare model predictions with temperature measurements. These resources support the modeling, identification, and control activities described here and broaden access to learners with a web browser, including those without a~physical kit \cite{tclab_apps}.

\section{Structure of relevant control engineering courses}

This section describes the education policy and curricula of control engineering courses taught at the University of West Bohemia in Pilsen (UWB). A survey of theoretical topics is presented to provide a context for the results presented in this paper.

The UWB adopts a structured study program model established through the Bologna Process started in 1999 and the European Higher Education Area formed in 2010. There is a~first undergraduate Bachelor's degree cycle taking 3 years and 180 credits based on defined learning outcomes and associated workload. A second graduate Master's degree cycle follows, with a standard duration of 2 years and 120 credits.

There are two main undergraduate study programs offered at the Department of Cybernetics of the Faculty of Applied Sciences involving control engineering as a main specialization topic:
\begin{enumerate}
    \item \textbf{Automatic Control and Robotics} focuses on theoretical knowledge and practical applications of cybernetics and informatics in the field of industrial automation and robotics. Students deal with advanced control systems, modeling and identification of systems and processes. A~key element is the integration of theoretical knowledge with practical skills in areas such as robotics engineering, systems diagnostics, and manufacturing process automation.
    \item \textbf{Artificial Intelligence and Automation} specialization focuses on the latest developments in areas such as machine learning, computer vision, and voice communication. Special emphasis is placed on deep neural networks, which are the basis of advanced technologies such as machine translation and generative artificial intelligence.
\end{enumerate}

While differing in the offered optional specialization subjects, both programs share a compulsory two-semester course \emph{Linear Systems 1+2}, providing a thorough introduction in the theory of dynamic systems in the first part, followed by a~review of classical and modern control theory topics in the second module.

As for the second cycle, two followup master degree programs exist, mirroring the two previously described first cycle counterpart's names and scope of study, while offering a series of more advanced courses in the respective domains of automation, robotics and artificial intelligence. There is an optional course \emph{Robust and multi-variable control} offered to last year master students, dealing with the control issues in complex multiple-input multiple-output (MIMO) dynamic systems. 

\begin{figure}[t]
\centering
\includegraphics[width=0.485\textwidth]{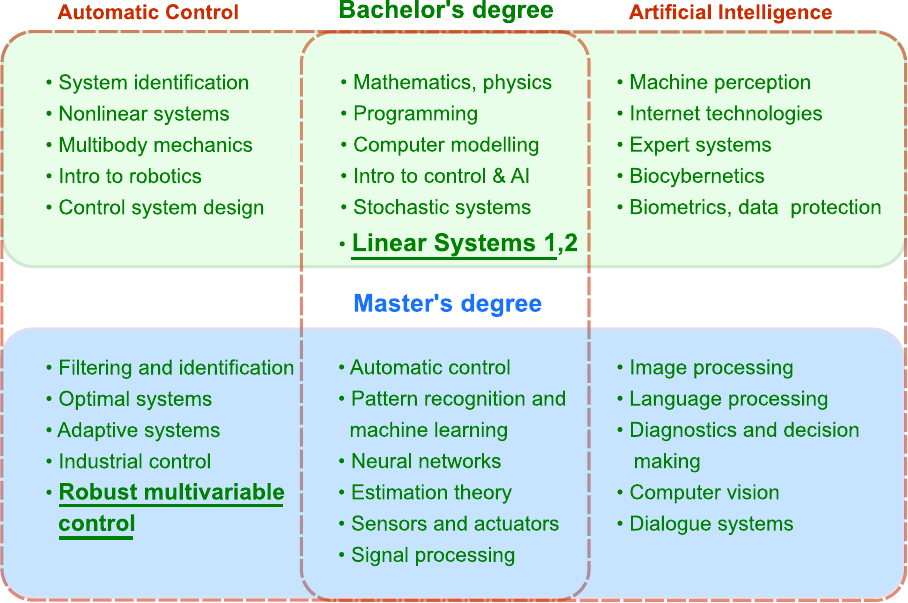}
\caption{Structure of Automatic Control and~Artificial Intelligence study programs offered at UWB. The left and right columns correspond to main specialization subjects of each program while the middle part denotes the common mandatory foundation courses.}
\label{fig_1}
\end{figure}

An overview of the study programs with the main relevant topics is given in Fig. \ref{fig_1}. Both the introductory \emph{Linear Systems~1} and the advanced \emph{Robust and Multivariable Control} courses have undergone a change in the practical semester projects, where the take-home labs described in this paper were employed to foster students' hands-on experience, as described in the following sections. 


\section{Introductory TCLab control project}
This section provides a study on how the take-home laboratory kits enhanced the introductory undergraduate \emph{Linear systems} course through a practical semester project. The goal for the students is to get acquainted with the setup and grasp the fundamental control engineering concepts in the simpler SISO scenario, allowing them to move to a~more complex MIMO setup later in the second follow-up graduate course. The project consists from three subsequent learning modules described below.

\subsection{First principle modeling}
The understanding of the temperature control system behavior can be established through the basic laws of thermodynamics. A simplifying assumption is made by modeling the heater elements as two point masses, with the equation of motion derived from the energy balance
\begin{align}
m c_p\frac{dT_1}{dt}=&UA(T_\infty-T_1)+\epsilon \sigma A \nonumber (T_\infty^4-T_1^4)\\
&+Q_{C12}+Q_{R12}+\alpha_1 Q_1\nonumber,\\
m c_p\frac{dT_2}{dt}=&UA(T_\infty-T_2)+\epsilon \sigma A \nonumber (T_\infty^4-T_2^4)\\
&-Q_{C12}-Q_{R12}+\alpha_2 Q_2,\label{eq:model}    
\end{align}
where $m[kg]$ is the mass of the heater assembly consisting from the transistor, thermal sensor and heatsink element, $c_p[J/kg\cdot K]$ is the heat capacity, $U[W/m^2\cdot K]$ is the overall heat transfer coefficient, $A[m^2]$ is the outer heatsink surface, $T_1,T_2,T_\infty[K]$ denote the temperatures of the two heaters and the ambient temperature, respectively, $\epsilon$ is the emissivity coefficient and $\sigma[W/m^2.K^4]$ is the Stefan-Bolzmann constant. 


Students receive the values of all the model parameters which are provided at the manufacturer's webpage. There are several follow-up questions in the project assignment designed to strengthen the understanding of the problem (with the hints to the correct answers):
\begin{itemize}
    \item What is the dominant source of heat losses - convection or radiation? (convection, see the relative magnitude of the scaling constants of the respective model terms)
    \item What simplifications of the plant model could be made to make it linear? (omit the radiation terms or perform local linearisation)
    \item What kind of modification of the physical setup could be done to minimize the losses? (insulation)
    \item What kind of modification of the physical setup could be done to minimize the mutual influence between the two heater units? (heaters distance, thermal shielding)
\end{itemize}

\textbf{ELOs:} Upon completion of this module, students should be able to:
\begin{enumerate}[label=M\arabic*)]
\item derive a simplified dynamic model of a thermal process from first-principles energy balances;
\item identify the physical meaning of the model parameters and assess their influence on system behavior;
\item propose justified model simplifications while discussing their effect on predictive accuracy.
\end{enumerate}

\subsection{Data-driven identification}
\begin{figure}[!t]
\centering
\includegraphics[width=0.485\textwidth]{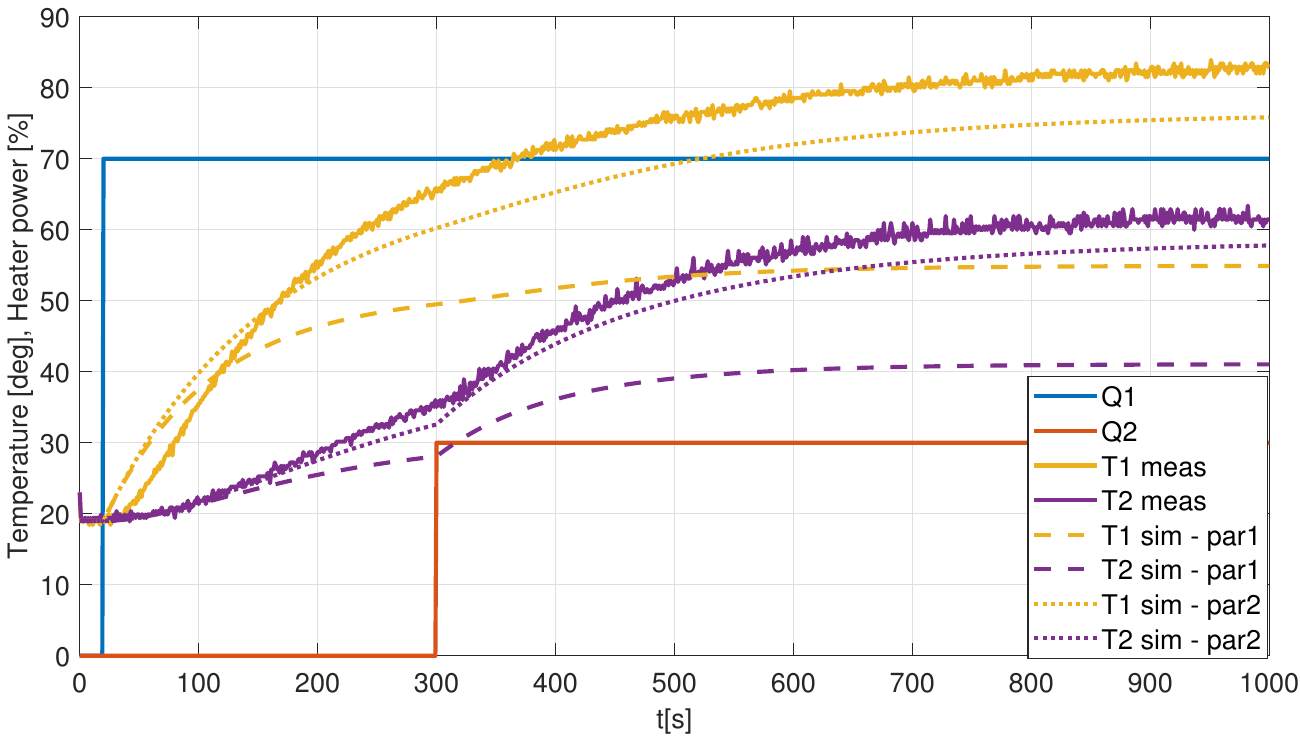}
\caption{Model validation - measured response of the physical setup compared to model predictions with two different parameter sets (top), closed-loop performance with the designed PI controller (bottom)}
\label{fig_model_vs_experiment}
\end{figure}

The initial results plotted by dashed lines in Fig. \ref{fig_model_vs_experiment} demonstrate that the derived model is capable of predicting the plant response \emph{qualitatively} (solid lines). However, there is a~significant discrepancy between the real and estimated steady-state temperatures. 

Students are asked to explain the obtained results:
\begin{itemize}
    \item What is the main source of the modeling errors - inaccurate parameter values or incorrect model structure?
    \item Can we improve the predictive fidelity of the model by tuning its parameters? If yes, which one shall we modify?
\end{itemize}




The next step is taken towards combining prior information about plant dynamics with measured data. Students employ numerical optimization or randomized sampling methods to find a better set of selected plant model parameters in (\ref{eq:model}), leading to improved model accuracy (Fig. \ref{fig_model_vs_experiment} dotted lines).


A \emph{black-box} approach is also studied by introducing first-order and second-order plus dead-time (FOPDT, SOPDT) structures, commonly used in the process control domain. The parameters of these control-oriented models are obtained again by fitting the experimental data using numerical optimization techniques.


\textbf{ELOs:} Upon completion of this module, students should be able to:
\begin{enumerate}[label=I\arabic*)]
\item validate dynamic plant model by comparing simulated and experimentally measured responses;
\item explain the advantages and limitations of white-box, gray-box, and black-box modeling approaches;
\item formulate an identification problem based on measured input–output data;
\item estimate unknown model parameters using numerical optimization or randomized sampling;
\item select a model structure appropriate for subsequent control design.
\end{enumerate}



\subsection{Model-based control design}
The previous learning modules provided a starting point for a model-based control design. A SISO control problem is established with the goal of controlling the temperature $T_1$ of the first heater element using the energy supplied by the corresponding transistor power $Q_1$. This can be done by forming a feedback loop and inserting a PI controller with the control law in the form of
\begin{equation}
Q_1(t)=K_pe(t)+K_i\int_0^t e(\tau)\,\mathrm{d}\tau;~e(t)\overset{\Delta}{=}T_1^{*}(t)-T_1(t),
\end{equation}
where $e(t)$ is the control error variable computed as the difference between the temperature setpoint $T_1^*$ and the actual value $T_1$ measured by the sensor and $K_p,K_i$ are the proportional and integral gains to be designed.

The role of proportional and integral parts is discussed at first, followed by trial-and-error experiments for finding a~suitable set of the PI gains using the physical setup. Students quickly realize that it is a rather difficult and time-consuming process, thanks to relatively slow dynamics and long time constants of the system. Improper setting of the controller parameters may lead to sluggish, oscillatory or unstable closed-loop response. This helps to understand the benefits of model-based simulation and control design. 

Students first learn to formulate basic performance requirements in time and frequency domain. The importance of closed-loop stability is also emphasized, due to the assumed imperfections in plant model and controller realization. Students then apply classical control design methods, starting with Evans' root locus method, followed by frequency-domain loopshaping or $H_\infty$ design to achieve sufficient stability margins \cite{brabec_hinf}. 

\textbf{ELOs:} Upon completion of this module, students should be able to:
\begin{enumerate}[label=S\arabic*)]
\item derive a set of suitable PI gains for the closed-loop temperature controller;
\item evaluate closed-loop stability and performance by means of numerical simulations;
\item implement the PI controller and validate its functionality using the physical setup.
\end{enumerate}

\section{Advanced TCLab control project}
The inherently multi-variable nature of the TClab kit is fully exploited in the follow-up Robust control course. Students first study the problem of \emph{interactions} arising from thermal coupling between the two heater elements and learn about their possible implications for control design.

The starting point is a derivation of a MIMO plant model obtained from (\ref{eq:model}) through a local linearization around a chosen equilibrium (steady state input power and output temperature) $Q_{1,2}(t)=Q_{10,20}+\Delta Q_{1,2}(t),~T_{1,2}(t)=T_{10,20}+\Delta T_{1,2}(t)$, where the linearized dynamics in the relative coordinates can then be expressed in the state space form
\begin{align}
    &\Delta \dot{\textbf{x}} = \textbf{A}\Delta\textbf{x} + \textbf{B}\Delta\textbf{u};~\Delta\textbf{x}\overset{\Delta}{=}\Delta\textbf{y}=[T_1-T_{10},T_2-T_{20}]^T,\nonumber\\
    &\Delta \textbf{y} = \textbf{C}\Delta\textbf{x} + \textbf{D}\Delta\textbf{u};~\Delta\textbf{u}\overset{\Delta}{=}[Q_1-Q_{10},Q_2-Q_{20}]^T,
\end{align}
with the real matrices $\textbf{A,B,C,D}$ obtained from the Taylor expansion of equations (\ref{eq:model}) in the vicinity of the chosen working point. Transfer function matrix can also be obtained to describe the input-output relation in the Laplace domain
\begin{align}
\textbf{Y}(s)=\textbf{P}(s)\textbf{U}(s);~\textbf{P}(s)=\frac{\begin{bmatrix}
        {k_1(s + z_1)} & & 
        k_2\\
        k_3& &
        {k_4(s + z_2)}
    \end{bmatrix}}{(s+p_1)(s
+p_2)},
\end{align}
where the parameters depend on the chosen equilibrium point. The properties of multi-variable poles and zeros and their corresponding directions are introduced. The role of the Smith-McMillan form in determining the minimal order of realization and in revealing zero-pole interactions and the internal structure of the plant dynamics.


The issue of \emph{input-output pairing} is explained, with two possible control loop interconnections $u_1\rightarrow y_1,u_2\rightarrow y_2$ and $u_1\rightarrow y_2,u_2\rightarrow y_1$. The concept of the relative gain array (RGA) is then studied as a framework for the interaction analysis and determination of suitable input-output pairing for decentralized control. The RGA analysis confirms a physical intuition behind the plant dynamics -- it is more straightforward to control each heater element's temperature using the corresponding transistor-sensor pair. Diagonal dominance and its connection with closed-loop stability through Gershgorin's theorem are also established.



The next part of the assignment deals with decoupling filter design. The goal is to precompensate plant inputs using a~static or dynamic decoupler to achieve partial or complete cancellation of interactions between the individual input-output channels.

In the next phase, students get acquainted with the centralized control design methods, leading to full multi-variable controllers. Students can select an arbitrary MIMO control method studied during the courses, ranging from Jordan form assignment via state feedback, LQG control, norm-based optimal control or Youla-Ku\v{c}era Q-parameterization. 

The last topic to learn is the introduction of uncertainty in the plant model, formulation and solution of a robust control problem. The goal is to address several imperfections of the first-principle model, such as sensor/actuator dynamics, varying behavior due to different operating points, and time-delay effects arising from gradual heat propagation. The methods of robust analysis and synthesis using (structured) singular values ($\sigma/\mu$ analysis ) are introduced \cite{skogestad2005multivariable}.  

\begin{figure*}[t]
\centering
\includegraphics[width=1\textwidth]{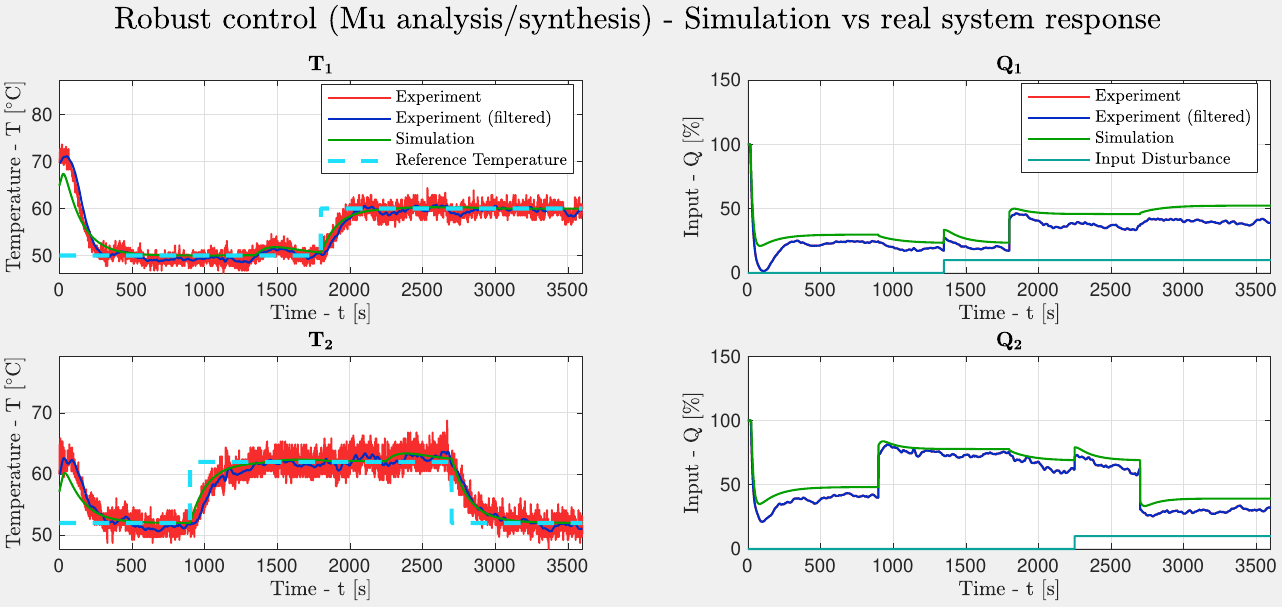}
\caption{MIMO control results - simulated and experimental results demonstrating temperature setpoint tracking, decoupling and disturbance rejection}
\label{fig_robust}
\vspace{-0.34cm}
\end{figure*}

Figure (\ref{fig_robust}) shows the results achieved with a properly implemented robust multi-variable controller. The left plot shows reference temperature tracking, while the right displays the plant inputs (actuator effort). It can be seen that perfect decoupling between the two outputs was achieved (changes in one desired temperature do not affect the second one). Input disturbances injected at $t=\{1400,2250\}$ are also handled well. Comparison with the simulated results (green lines) also shows a good match, advocating the use of model-based control design.

\textbf{ELOs:} Upon completion of this module, students should be able to:
\begin{enumerate}[label=C\arabic*)]
\item derive linearized multi-variable plant models;
\item analyse dynamic interactions, identify suitable input-output pairing for decentralized control;
\item design static and dynamic decoupling filters;
\item formulate and solve multi-variable control design problem using various approaches (modal control, quadratic-optimal control, norm-based $H_2/H_{\infty}$ synthesis, Internal model control...);
\item describe plant model uncertainty and integrate it into a robust control design problem, solution via mixed-sensitivity $H_\infty$ control and $\mu$ synthesis, implement robust controller and validate it using the physical setup.
\end{enumerate}

\newcolumntype{L}{>{\RaggedRight\arraybackslash}X}

\begin{table*}[!t]
\centering
\caption{Mapping of competency domains to expected learning outcomes, learning activities, and evidence of achievement (RQ3)}
\label{tab:competency-domains}
\small
\setlength{\tabcolsep}{3pt}
\renewcommand{\arraystretch}{1.20}

\begin{tabularx}{\textwidth}{@{}L L L L@{}}
\toprule
\textbf{Control engineering competency domain} &
\textbf{Related ELOs} &
\textbf{Main learning activities} &
\textbf{Evidence of achievement} \\
\midrule

Physical Modelling and System Understanding
&
ELO-M1--M3
&
First-principles modelling, formulation of energy balances, implementation of the dynamic model, and analysis of physical parameters.
&
Model equations, simulation results, parameter interpretation, and a technically justified discussion of modelling assumptions.
\\
\addlinespace

System Identification and Model Validation
&
ELO-I1--I5
&
Parameter estimation, analysis of measured input--output data, comparison of simulated and experimental responses, evaluation of model fidelity.
&
Validation plots, quantitative error measures, identified parameters, and an assessment of model-structure and parameter uncertainty.
\\
\addlinespace

SISO Control Design and Implementation
&
ELO-S1--S3
&
Formulation of a single-input single-output control problem, PID controller tuning, closed-loop simulation, and implementation on the physical setup.
&
Controller parameters, simulation results, experimentally measured closed-loop responses, and an interpretation of performance metrics.
\\
\addlinespace

MIMO Analysis and Control Design
&
ELO-C1--C5
&
Analysis of input--output interactions, formulation of a multivariable control problem, MIMO controller design, and experimental implementation.
&
Interaction analysis, controller design documentation, experimental data, closed-loop performance evaluation, and justification of design decisions.
\\
\addlinespace

Experimental Evaluation and Engineering Practice
&
ELO-I1,I4,S3,C5
&
Execution of hardware experiments, evaluation of real-system behavior, troubleshooting, and technical reporting.
&
Experimental records, comparison of measured and simulated responses, explanation of discrepancies, and coherent engineering conclusions in project documentation.
\\

\bottomrule
\end{tabularx}
\vspace{1cm}
\end{table*}

\section{Evaluation of student learning and course performance}

The remainder of this section is organized around the three research questions formulated in Section I. RQ1 is addressed using the student feedback survey (Section VI-A), RQ2 is examined through the analysis of midterm test scores and final exam grades (Section VI-B), and RQ3 is discussed by mapping the project modules and ELOs onto the five competency domains (Table I). 

The summary Table I organizes the expected learning outcomes into five broader control engineering competency domains targeted in the individual project modules. Each domain is linked to the corresponding ELO identifiers used throughout the course description, thereby providing a concise overview of the learning activities and the evidence used to assess their achievement. This structure connects the technical tasks performed by the students with the intended educational outcomes and their assessment.

Because the available historical data did not provide direct assessments for each individual ELO, the effect of the curricular change was evaluated using indirect indicators, including student feedback, midterm test scores, and final exam grades.

\subsection{Student feedback survey}
We provide results of a questionnaire filled by an experimental group B described below, which completed the updated semester projects with the use of the take-home lab kits. The goal was to evaluate their hands-on experience and compare it with a previous introductory systems and control course, where only numerical simulations were used instead of experiments with physical hardware.

The students rated the validity of the following statements:
\begin{enumerate}
    \item The assigned temperature control project using TCLab kit was interesting.
    \item I find experiments with real hardware more useful for learning than purely numerical offline simulations, e.g. in Matlab/Simulink.
    \item The semester project helped me better understand the model-based development cycle in control engineering problems (modeling-design-simulation-implementation-experiments).
    \item The semester project helped me better understand the theory from the lectures.
\end{enumerate}

\begin{figure}[t]
\centering
\includegraphics[width=0.485\textwidth]{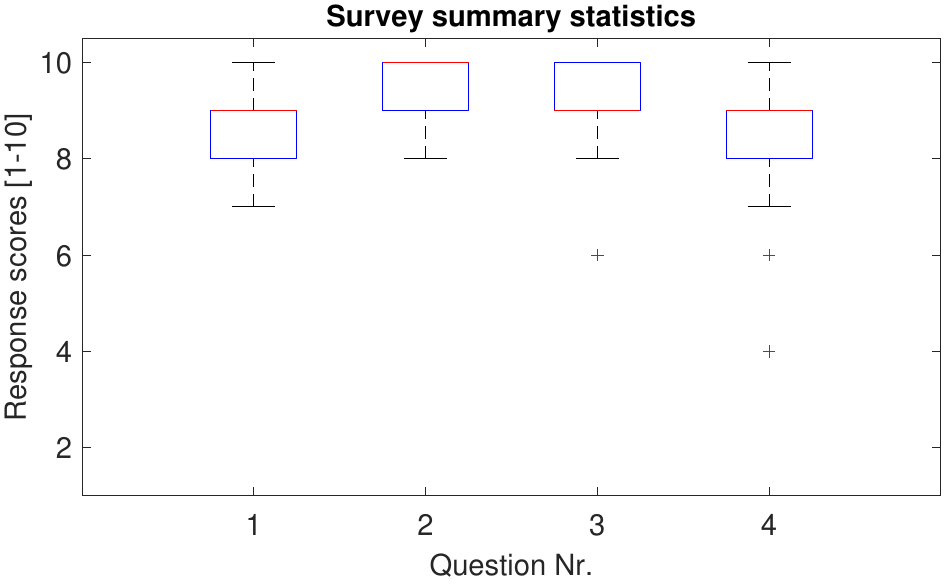}
\caption{Student feedback survey - boxplots summarizing the statistics of the received questionnaires. The red line denotes the median of each dataset, blue rectangles show the inter-quartile range, the whiskers extend to the minimum and maximum values and red markers designate outliers}
\label{fig_survey}
\end{figure}


The answers were given on the scale of 1-10, ranging from \emph{strongly disagree} to \emph{completely agree}. The results are summarized in the boxplots in Fig. \ref{fig_survey}. The survey results provide evidence relevant to RQ1, indicating high acceptance of the new semester project and subjectively positive impacts on learning.

\subsection{Score-based Evaluation of Course Performance}
The learning outcomes were also evaluated by means of the scores achieved by students during midterm written tests and final grades received from the course exam from the period of last 10 years.

The score-based evaluation addresses RQ2 by comparing two distinct group of students:

\begin{itemize}
    \item \emph{Group A} - control group containing 281 students which completed the above described control engineering courses during the years 2016-2023 prior to the proposed changes in the semester projects.
    \item \emph{Group B} - the experimental group of 57 students undergoing the same courses in the last two academic years starting in 2024-2025, after the introduction of the modified projects employing the take-home lab kits.
    
\end{itemize}
Apart from the semester project, there were no other changes in the course curricula, teaching methods, materials, teachers or exam content.

At first we compare the scores received from two midterm written tests, which are mandatory to pass the course. Two tests held in two-thirds and before the end of the semester evaluate practical skills such as calculations of time and frequency-domain responses of systems, manipulation of various forms of models, analysis of basic system properties such as stability, controllability or observability.

\begin{figure}[t]
\centering
\includegraphics[width=0.485\textwidth]{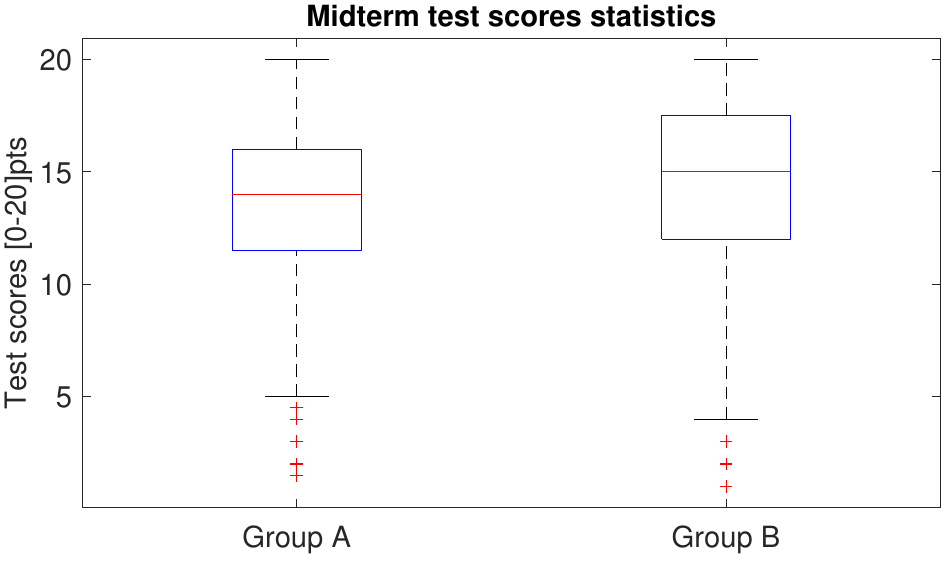}
\caption{Midterm test scores statistics - boxplots summarizing the scores achieved by the studied student groups in the midterm tests}
\label{fig_test_scores}
\end{figure}

Figure \ref{fig_test_scores} shows the boxplots computed from the datasets containing the test scores for the two groups. The scores range from 0-20 points, with a minimum 10 point treshold value to pass the test. We can see a slight improvement of the Group B results, noticeable in the higher median value (15 pts versus 14) and shifted interquartile range (12-17.5 vs 11.5-16). 

We applied the Mann–Whitney U test, a nonparametric procedure for assessing statistically important differences between the two  groups. This test was selected because the score distributions did not meet the assumptions required for a~parametric independent-samples t-test, in particular normality, and because the cohort sizes were unequal. The null hypothesis $H_0$ posited that both cohorts were drawn from populations with identical distributions, whereas the alternative hypothesis $H_1$ stated that the distributions differ. 

 The test yielded a p-value of $p=0.048$, which is below the conventional significance threshold of 0.05. Accordingly, the null hypothesis was rejected, indicating a statistically significant difference between the two cohorts. However, the effect size was relatively small. The rank-based effect size computed as $r=|z|/\sqrt{N}=0.115$, where $z$ is the test z-value and $N$ is the number of observations, indicates a weaker association between group membership and test scores, and Cliff’s delta counting the dominance of every pair in the two datasets $\delta=\frac{\#x_i>y_j-\#x_i<y_j}{n_A\cdot n_B}=-0.141$  likewise suggests a small practical difference between the cohorts, with the negative sign indicating higher scores in the post-intervention group. 
 
 Taken together, these metrics show that although the difference was statistically significant, its magnitude was modest. A possible interpretation of this phenomenon is that students naturally concentrate most of their effort spent on the project at the semester end, as the deadline for submission approaches. 
 

\begin{figure}[t]
\centering
\includegraphics[width=0.485\textwidth]{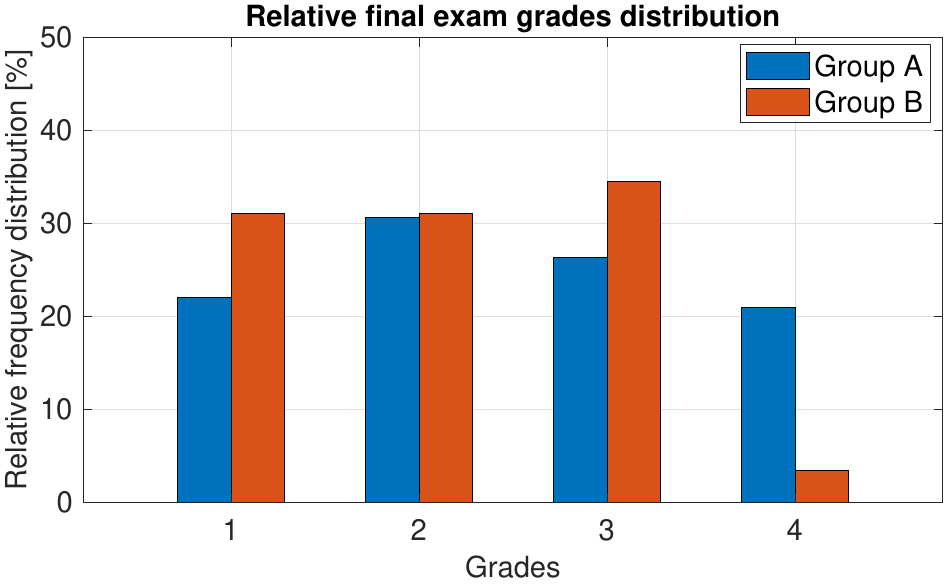}
\caption{Final exam grades statistics - barchart showing the relative frequency distribution}
\label{fig_grades}
\end{figure}

 On the other hand, we revealed a more significant shift in the overall course grades received during final examination. Figure \ref{fig_grades} shows the bar charts counting the frequency distribution of the received grades on the scale of \emph{1 - excellent, 2 - very good, 3 - satisfactory, 4 - did not pass}. We can see a~notable difference in the grades distribution. Group A contains a~relatively high proportion of 21 percent of students who did not pass the course (mark 4). Approximately one half of this group gave up during the semester and did not come to the final exam at all while the other half failed to meet the minimum course requirements. The failure rate dropped from 21 to 3.4 percent in Group B. The portion of excellent students with the best grade increased from 22 to 31 percent, while the lowest score group with ranking 3 increased from 26 to 34 percent. The mean score in the pre-intervention Group A was $\bar{x}_A=2.46$ with standard deviation $s_A=1.06$, whereas the post-intervention Group B achieved $\bar{x}_B=2.1$ and $s_B=0.89$, indicating an overall shift to better grades.

 The Mann-Whitney test was performed again on the grades dataset to assess the statistical significance and effect size of the observed changes. The test yielded a p-value of $p=0.0224<0.05$, indicating a statistically significant change in the distribution of grades. Cliff's delta of $\delta = 0.184$ shows the dominance of the first dataset, indicating overall worse grades of Group A, together with a rank-based score of $r = 0.124$, revealing a relatively larger effect size compared to the interim testing results.

The observed significant decrease in the failure rate, together with the results of the self-assessment survey, indicates that the hands-on experience provided by the take-home labs has a measurable effect on students’ motivation to complete the course and pass the final exam. 


\section{Conclusion}

We have presented a sequence of basic and advanced semester projects based on take-home TCLab kits that can be integrated into control engineering courses at technical universities. The projects are embedded in a project-based learning and competency-based education framework and target five control engineering competency domains through explicitly formulated expected learning outcomes.

Evaluation using a student feedback survey, midterm test scores, and final exam grades revealed a statistically significant difference between the pre- and post-intervention cohorts, albeit with small effect sizes. The findings provide affirmative answers to the three research questions from Section~I: (RQ1) students report improved understanding and engagement; (RQ2) there is a statistically significant, albeit modest, improvement in exam performance; and (RQ3) the proposed modules systematically support the development of the targeted competency domains. Nevertheless, the practical impact of the curriculum change should be interpreted cautiously and considered together with other contextual factors such as cohort composition, assessment difficulty, and sample size.

Beyond exam performance, the curriculum change appears to have enhanced learning outcomes that are not easily captured by grades alone, such as conceptual understanding, confidence in applying theory to real problems, and overall engagement. Future work should examine additional cohorts and include more direct measures of technical competency development, for example through rubric-based assessment or concept inventories. The proposed take-home lab projects can serve as a template for similar interventions in other control engineering courses and institutions.

\section*{Acknowledgments}
This work was created with the support of the project "R\&D of Technologies for Advanced Digitization in the Pilsen Metropolitan Area (DigiTech)" No.: CZ.02.01.01/00/23\_021/0008436 co-financed by the European Union.



 
%

\bibliographystyle{IEEEtran}

\bibliography{references}


 




\vfill

\end{document}